\documentclass[cameraready]{Interspeech}
\usepackage{booktabs} 
\usepackage{multirow} 
\usepackage{graphicx} 
\usepackage{amssymb}  
\usepackage{array}   
\usepackage{hyperref}
\usepackage{multirow}
\usepackage{colortbl,xcolor}
\usepackage[normalem]{ulem}
\AtBeginDocument{
  \setlength{\abovedisplayskip}{2pt plus 1pt minus 1pt}
  \setlength{\belowdisplayskip}{2pt plus 1pt minus 1pt}
  \setlength{\abovedisplayshortskip}{0pt plus 1pt}
  \setlength{\belowdisplayshortskip}{2pt plus 1pt minus 1pt}
}

\title{Text-Independent Speaker Verification Using Discrete Audio Tokens}

\author[affiliation={1}, orcid=0009-0000-9252-7579]{Zheng}{Liang}
\author[affiliation={1}, orcid=0009-0007-9654-9223]{Junjie}{Li}
\author[affiliation={1},orcid=0000-0001-9133-3000,correspondingauthor]{Kong Aik}{Lee}

\address{
  $^1$ The Hong Kong Polytechnic University, Hong Kong SAR, China
}

\email{zheng.liang@connect.polyu.hk, junjie98.li@connect.polyu.hk, kong-aik.lee@polyu.edu.hk}


\usepackage{comment}

\begin{document}

\maketitle

\begin{abstract}

Neural audio codecs (NACs) enable efficient audio compression and have achieved success in downstream tasks such as speech synthesis. However, their discrete representations consistently underperform traditional spectral features in automatic speaker verification (ASV). We empirically demonstrate that speaker cues are implicitly preserved in discrete tokens but remain underutilized by conventional ASV training paradigms. To address this, we propose a Cross-Feature Knowledge Distillation (CFKD) framework. By guiding the codec-based student to mimic the embedding space of a strong Fbank-based teacher, CFKD provides structured supervision for effective utilization of speaker information in tokens. Experiments on the VoxCeleb benchmarks show that CFKD substantially improves the ASV performance of codec-based systems, allowing them to approach the accuracy of Fbank-based teacher models and highlighting the potential of discrete audio tokens for diverse speech tasks.

\end{abstract}
\section{Introduction}

Text-independent Automatic Speaker Verification (ASV) aims to authenticate a speaker’s identity regardless of phonetic content, requiring the extraction of robust and speaker-specific traits from unconstrained speech signals~\cite{wang2024overview}. State-of-the-art systems predominantly rely on handcrafted acoustic features, such as Mel-frequency cepstral coefficients or filterbank energies (Fbanks), which are carefully designed to capture perceptually relevant spectral characteristics~\cite{desplanques2020ecapa, snyder2018x, wang2023cam++, liu2024golden}. These spectral representations have long served as the standard front-end for ASV and a wide range of speech processing tasks~\cite{mehrish2023review, hinton2012deep,eyben2015geneva}.

In recent years, the speech processing research has witnessed rapid advances in neural audio codecs (NACs)~\cite{kumar2023high,zeghidour2021soundstream, defossezhigh}. By compressing raw waveforms into sequences of hierarchical discrete tokens through residual vector quantization, NACs enable high-fidelity audio reconstruction at extremely low bitrates~\cite{kumar2023high,zeghidour2021soundstream, defossezhigh}. Owing to their compactness and generative capability, these discrete representations have become foundational building blocks for large audio models, speech synthesis, and speech language modeling~\cite{chen2025neural, borsos2023audiolm}.

Despite their remarkable success in generative applications, NAC tokens exhibit a pronounced performance degradation when directly employed for recognition tasks, particularly ASV~\cite{lin2025codec, mousavi2024dasb,puvvada2024discrete}. This degradation presents a fundamental contradiction: If tokens preserve sufficient information to faithfully reconstruct speech timbre, prosody, and other core components of speaker identity, they should be adequate for speaker discrimination.

To investigate this discrepancy, we conduct a systematic analysis of ASV performance using three forms of input: raw waveforms, reconstructed waveforms generated by the codec decoder, and discrete codec tokens. Our results indicate that direct token-based systems exhibit a substantial performance drop, which is in line with prior works~\cite{lin2025codec, mousavi2024dasb,puvvada2024discrete}, whereas the system operating on reconstructed waveforms has a slight degradation relative to those using original waveforms. This comparison suggests that NACs likely retain most speaker-related information during compression. Instead, the performance gap may primarily arise from the challenges of effectively exploiting highly compressed and discrete latent representations under conventional ASV training paradigms.  

To better exploit the speaker information encoded in tokens, we propose a Cross-Feature Knowledge Distillation (CFKD) framework. In CFKD, a teacher model with rich, continuous spectral representations provides dense supervision to a token-based student model, guiding it to capture latent speaker-discriminative structures that are otherwise difficult to learn under vanilla cross-entropy-based training. By mimicking the embedding space of a strong Fbank-based teacher, the student model is encouraged to elicit and leverage the inherent speaker cues preserved in neural codec tokens, enabling more effective utilization of discrete representations for ASV.

\begin{figure}[t]
 \centering
 \includegraphics[width=\linewidth]{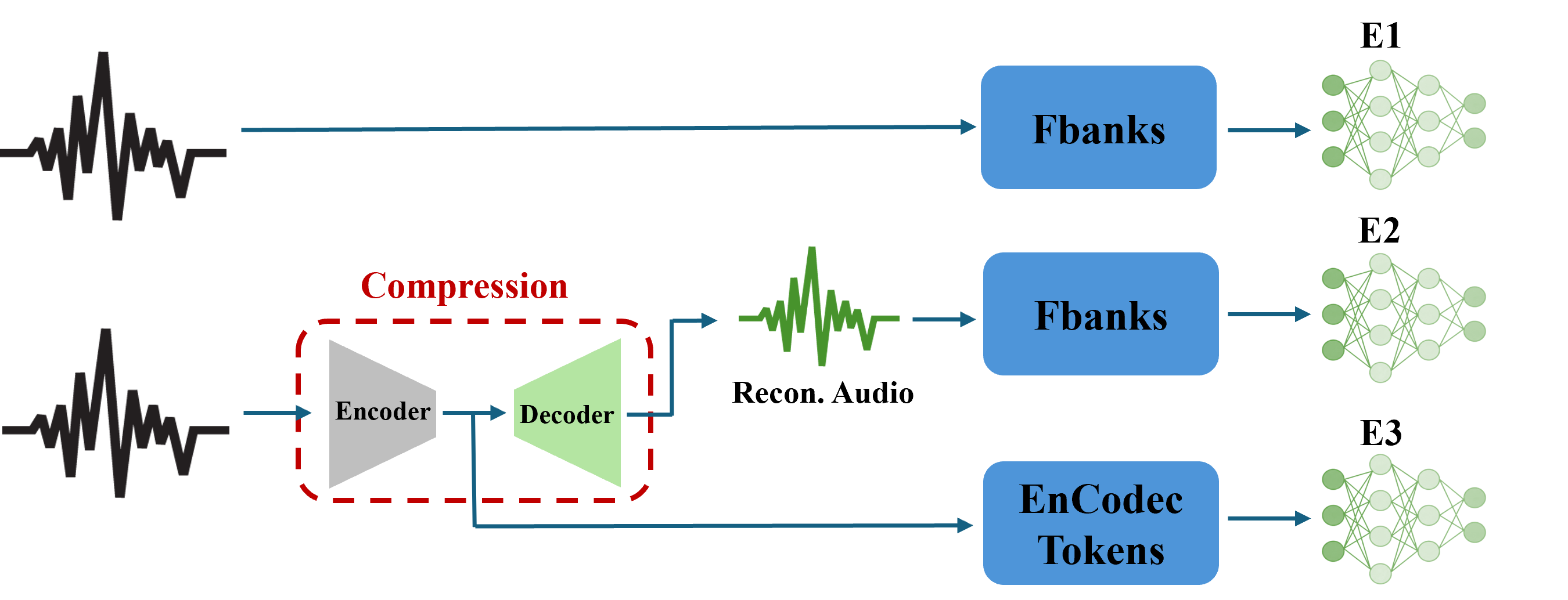}
 \caption{Schematic of the diagnostic benchmarks. Top (E1): The standard baseline using original Fbanks. Middle (E2): The reconstructed baseline using Fbanks extracted from the decoded audio. Bottom (E3): The ASV system trained directly on EnCodec tokens.}
 \label{fig:paradox_setup}
\end{figure}

\begin{figure*}[t]
 \centering
 \includegraphics[width=1\linewidth]{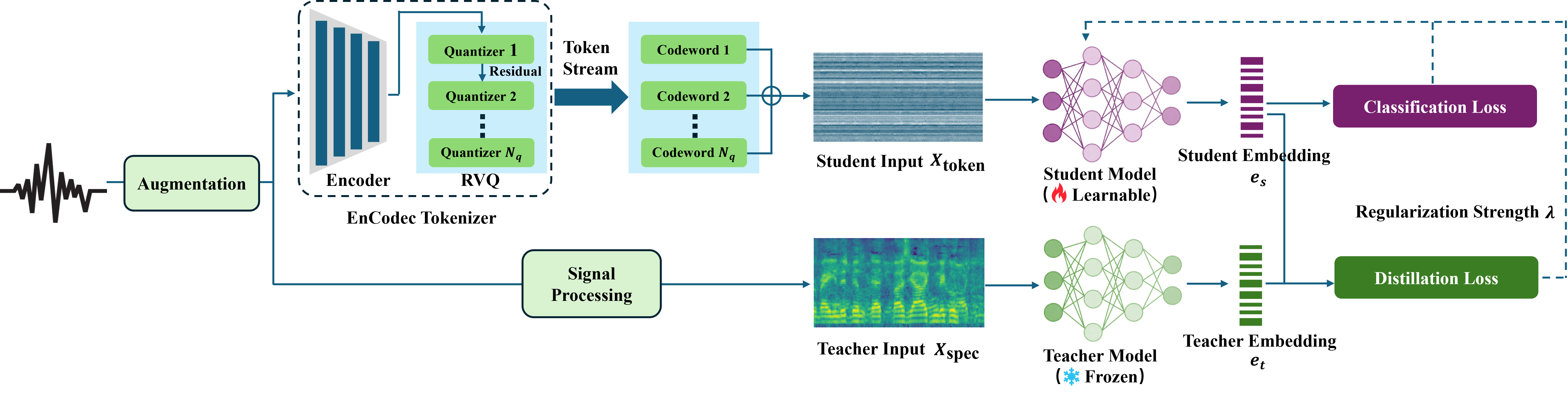} 
 \caption{Schematic illustration of the proposed CFKD framework. The architecture processes raw audio through dual streams: a continuous Fbank-based teacher and a discrete token-based student. The student input is derived by summing the codebook embeddings from the hierarchical quantizers of the neural codec.}
 \label{fig:framework}
\end{figure*} 

\section{From Continuous Spectrograms to Discrete Codec Tokens}
Modern ASV systems typically function as a sequence-to-vector mapping framework, where a deep neural backbone transforms input acoustic sequences into a discriminative latent speaker space~\cite{wang2024overview}. These systems predominantly rely on continuous spectral features like Fbanks~\cite{desplanques2020ecapa, snyder2018x, wang2023cam++, liu2024golden}. Let $\mathbf{X}_{\text{spec}} \in \mathbb{R}^{T \times F}$ denote the input Fbank features, where $T$ represents the temporal sequence length and $F$ corresponds to the frequency bins. These features are physically explicit, exhibiting strong local correlations in both time and frequency domains.

{
In contrast to continuous spectral representations, NACs introduce a learned discrete representation of speech through neural compression. By transforming raw waveforms into compact quantized tokens, NACs significantly reduce bitrate while preserving perceptual quality~\cite{kumar2023high,zeghidour2021soundstream, defossezhigh}. This discrete formulation enables flexible control over compression efficiency and facilitates scalable storage and transmission. 
Taking EnCodec~\cite{defossezhigh}, a representative streaming neural codec, as an example, its architecture comprises a convolutional encoder, a residual vector quantizer (RVQ), and a decoder. The encoder downsamples the waveform into a low-frame-rate latent sequence, which the RVQ discretizes into $N_q$ hierarchical quantization layers. The decoder reconstructs the signal from these quantized representations. The number of quantizers $N_q$ controls the trade-off between bitrate and reconstruction fidelity.}

{
Formally, let $\mathbf{Z} \in \{1, \dots, K\}^{T \times N_q}$ denote the quantized token matrix, where $K$ is the codebook size. Each token is mapped to its corresponding codebook embedding $\mathcal{C}_q \in \mathbb{R}^{K \times D}$ for the $q$-th quantizer, with $D$ being the embedding dimension. To obtain a dense speech representation suitable for the speaker encoder~\cite{lin2025codec, mousavi2024dasb,puvvada2024discrete}, we fuse information across all quantization layers by summing the embeddings along the quantizer dimension:
\begin{equation}
  \mathbf{X}_{\text{token}} \in \mathbb{R}^{T \times D} = \sum_{q=1}^{N_q} \text{Embed}(Z_{:,q}; \mathcal{C}_q) .
\end{equation}
This simple summation effectively reconstructs an approximate latent frame from the hierarchical residual components for downstream token-based models.}

\section{Methodology}
\subsection{Motivation: The Information Accessibility Gap}
\label{ssec:motivation}

{Before introducing our proposed framework, we conduct a diagnostic study to identify the sources of performance degradation in token-based ASV. Specifically, we compare three setups using the same ECAPA-TDNN backbone, as illustrated in Figure~\ref{fig:paradox_setup}. }

\begin{table}[htbp]
 \caption{Diagnostic analysis of the performance gap between tokens and Fbanks. Models are trained on Vox1 dev set and evaluated on Vox1-O test set.}
 \label{tab:paradox}
 \centering
 \begin{tabular}{llcc}
  \toprule
  ID & System Setup & EER (\%) & minDCF \\
  \midrule
  E1 & Original Fbanks   & 2.21 & 0.236 \\
  E2 & Reconstructed Fbanks  & 2.57 & 0.280 \\
  E3 & EnCodec Tokens   & 3.38 & 0.366 \\
  \bottomrule
 \end{tabular}
\end{table}

The results in Table~\ref{tab:paradox} reveal a clear trend. Comparing E1 and E2, the EER increases slightly (2.21\% $\to$ 2.57\%), indicating that the majority of speaker-discriminative cues are preserved during neural compression. In contrast, E3, which directly uses the discrete tokens produced by the EnCodec encoder, suffers a substantially higher EER (3.38\%). Although E3 is ultimately derived from the same underlying information as E2, the discrete and highly compressed nature of the tokens makes it more challenging for conventional ASV backbones to extract speaker-relevant information.

{This suggests that \textit{the performance drop in token-based ASV is not primarily due to the loss of speaker information, but rather because conventional training paradigms struggle to extract it from the highly compressed discrete tokens}.}

\begin{table*}[htbp]
 \caption{Performance comparison of the proposed CFKD framework across different backbones. Note that to enforce architectural symmetry, each student is distilled from a teacher sharing the identical backbone. Teacher models use Fbanks, while all Student models use EnCodec tokens. Results in gray denote the baseline system. The best performance is shown in \textbf{bold}. $\Delta$ means relative improvement compared to the baseline system. Models are trained on Vox1 dev set and evaluated on Vox1-O test set.} 
 \label{tab:main_results_rel}
 \centering
 \small
 \setlength{\tabcolsep}{5pt}
 \begin{tabular}{l ccc c ccc cc} 
  \toprule
  \multirow{2}{*}{Setting ($\lambda$)} & 
  \multicolumn{4}{c}{ECAPA-TDNN1024} & & 
  \multicolumn{4}{c}{ResNet34} \\
  \cmidrule(lr){2-5} \cmidrule(lr){7-10}
   & ID & EER (\%) & minDCF & $\Delta$(\%) & & ID & EER (\%) & minDCF & $\Delta$(\%) \\
  \midrule
  Teacher (Fbank) & E1 & 2.21 & 0.236 &  - & & E8 & 2.58 & 0.275 &  - \\
  \midrule
  \cellcolor{gray!20}Student ($\lambda=0$, Naive) & \cellcolor{gray!20}E3 & \cellcolor{gray!20}3.38 & \cellcolor{gray!20}0.366 & \cellcolor{gray!20} Benchmark & \cellcolor{gray!20} & \cellcolor{gray!20}E9 & \cellcolor{gray!20}7.55 & \cellcolor{gray!20}0.669 & \cellcolor{gray!20} Benchmark \\
  Student ($\lambda=10$)    & E4 & 2.51 & 0.290 & 23.3 & & E10 & 5.07 & 0.491 & 29.8 \\
  Student ($\lambda=20$)    & E5 & \textbf{2.25} & 0.265 & 30.6 & & E11 & 4.30 & 0.409 & 41.0 \\
  Student ($\lambda=40$)    & E6 & \textbf{2.25} & \textbf{0.231} & \textbf{35.3} & & E12 & \textbf{4.03} & \textbf{0.408} & \textbf{42.9} \\
  Student ($\lambda=80$)    & E7 & 2.30 & 0.256 & 31.1  & & E13 & 4.23 & 0.428 & 40.0 \\
  \bottomrule
 \end{tabular}
\end{table*}

\subsection{Cross-Feature Knowledge Distillation (CFKD)}

\subsubsection{Embedding-level Dense Supervision}

{Current ASV backbones are typically trained using cross-entropy (CE) losses, where hard labels treat all non-target classes as equally incorrect. This sparse supervision provides limited guidance for navigating the highly compressed and discrete latent space of neural codec tokens~\cite{hinton2015distilling}.}

{To address this limitation, we propose the CFKD framework as shown in Figure~\ref{fig:framework}, where supervision is imposed at the speaker embedding level. Specifically, the teacher extracts embeddings from Fbank features, while the student derives embeddings from discrete codec tokens. Both models share the same backbone architecture, ensuring that performance differences stem from feature representations rather than structural discrepancies.}

{By leveraging a pre-trained Fbank-based teacher, CFKD provides dense supervision through soft target embeddings, enabling the token-based student to capture speaker-discriminative cues that are difficult to learn under conventional CE training. Unlike logits-based distillation designed for classification tasks, embedding-level distillation directly aligns the student’s speaker embedding space with the well-structured geometry of the teacher~\cite{wang2019knowledge, gan25_interspeech,jung2019short}. }

\subsubsection{Geometric Alignment Objective}
Formally, we define the speaker encoding process as a mapping $\Phi: \mathcal{X} \to \mathcal{Z}$, where $\mathcal{Z} \subset \mathbb{R}^d$ is a hyperspherical embedding space. Let $\Phi_\text{T}$ be the pre-trained Fbank-based teacher and $\Phi_\text{S}$ be the token-based student. For a given utterance, we obtain the teacher embedding $\mathbf{e}_\text{t} = \Phi_\text{T}(\bf{X}_{\text{spec}})$ and the target embedding $\mathbf{e}_\text{s} = \Phi_\text{S}(\bf{X}_{\text{token}})$.

Since modern ASV objectives rely on angular discriminability~\cite{wang2018additive, deng2019arcface}, we employ a Cosine Similarity Loss to enforce directional alignment. This ensures the student captures the semantic orientation of the teacher's template embedding, effectively transferring the structural knowledge of the teacher model:
\begin{align}
      \mathcal{L}_\text{KD} &= 1 - \cos(\mathbf{e}_\text{s},  \mathbf{e}_\text{t}), \\
      &= 1 - \frac{\mathbf{e}_\text{s}^{\top} \mathbf{e}_\text{t}}{\lVert \mathbf{e}_\text{s} \rVert \lVert \mathbf{e}_\text{t} \rVert}. 
\end{align}
The total training objective combines the standard classification loss $\mathcal{L}_{\text{CLS}}$ with this distillation loss $\mathcal{L}_\text{KD}$:
\begin{equation}
  \mathcal{L}_{\text{TOTAL}} = \mathcal{L}_{\text{CLS}} + \lambda \mathcal{L}_\text{KD}, 
\end{equation}
where $\lambda$ controls the regularization strength.

\section{Experimental Settings}
\subsection{Audio Frontends and Backends} 
For the spectral baseline, we extract 80-dimensional log Fbanks directly from the original 16 kHz waveforms.

In contrast, for the token-based systems, we utilize the official pre-trained 24 kHz EnCodec model~\cite{defossezhigh} as the tokenizer. Consequently, all audio recordings are upsampled to 24 kHz prior to tokenization to align with the pre-trained model's specifications. We employ the full configuration with 32 residual vector quantizers. The discrete tokens are aggregated via summation and subsequently mapped to an 80-dimensional space using a learnable linear projection. This alignment ensures that the input dimensionality remains consistent across input features, guaranteeing parameter consistency for the backends. 

We evaluate two architectures: ECAPA-TDNN ($C=1024$)~\cite{desplanques2020ecapa} and ResNet34 ($C=32$)~\cite{he2016deep}, with 14.65M and 6.63M parameters respectively. Both backends produce 192-dimensional speaker embeddings.
\subsection{Training and Evaluation Protocol} 

Unless otherwise stated, models are trained on the VoxCeleb1 development set~\cite{nagrani2017voxceleb}, comprising 1,211 speakers. For these experiments, we adopt a standard data augmentation chain including speed perturbation~\cite{ko2015audio}, and additive noise/reverberation using the MUSAN~\cite{snyder2015musan} and RIR~\cite{ko2017study} datasets. The training process spans 100 epochs with a batch size of 256 and 2-second segments. We utilize the Adam optimizer with an initial learning rate of $1.2 \times 10^{-3}$, weight decay of $2 \times 10^{-5}$, and a step decay factor of 0.97. The loss function employed is AAM-Softmax~\cite{deng2019arcface} with $m=0.2$ and $s=32$.

As a specific exception, to evaluate the proposed method on a larger scale, we trained a separate model on the VoxCeleb2 dataset with 5994 speakers~\cite{chung2018voxceleb2} . For this experiment, we followed the default training recipe provided by the Wespeaker framework~\cite{wang2023wespeaker}. 

Performance is reported using EER and minDCF ($P_{target}=0.01$, $C_{miss}=C_{fa}=1$).

\section{Experimental Results}

\subsection{The Effectiveness of Distillation}
\begin{table*}[t]
\centering
\caption{Performance comparison on Vox1 test sets across various bitrates. Results are reported in \textbf{EER (\%) / minDCF} format. While both methods used the ECAPA-TDNN1024 backbone, our model ($\lambda=40$) is trained on Vox2 with the pre-trained teacher from Wespeaker, and codec-ASV is trained on Vox1+Vox2, so their results are reported on Vox1-O only. (*) indicates the result was reported at 18 kbps.}
\label{tab:bandwidth_comparison_final_v2}
\resizebox{0.8\linewidth}{!}{ 
\begin{tabular}{l l c c c c c |c}
\toprule
\multirow{2}{*}{Test Set} & \multirow{2}{*}{Method} & \multicolumn{5}{c|}{Bitrate} & \multirow{2}{*}{Teacher} \\
\cmidrule(lr){3-7}
& & 1.5 kbps & 3 kbps & 6 kbps & 12 kbps & 24 kbps & \\
\midrule
\multirow{2}{*}{Vox1-O} 
& Codec-ASV (M5)~\cite{lin2025codec}      & 19.32 / 0.967 & 8.87 / 0.819 & 3.08 / 0.396 & --           & 2.08 / 0.307\textsuperscript{*} &\multirow{2}{*}{0.86 / 0.090} \\
& \textbf{Ours} & \textbf{9.24 / 0.736} & \textbf{3.06 / 0.345} & \textbf{1.53 / 0.196} & \textbf{1.15 / 0.148} & \textbf{1.05 / 0.118} & \\
\midrule
Vox1-E 
& Ours & 9.72 / 0.787 & 3.47 / 0.386 & 1.80 / 0.208 & 1.30 / 0.150 & 1.20 / 0.139 & 1.07 / 0.117 \\
\midrule
Vox1-H 
& Ours & 18.37 / 0.907 & 6.96 / 0.543 & 3.56 / 0.325 & 2.62 / 0.250 & 2.39 / 0.230 & 2.06 / 0.205 \\ 
\bottomrule
\end{tabular}
}
\end{table*}

{In this section, we evaluate the effectiveness of the proposed CFKD training framework across different backbone architectures, including ECAPA-TDNN and ResNet, in terms of EER and minDCF, as reported in Table~\ref{tab:main_results_rel}. The results show that, under various settings of the distillation weight $\lambda$, all models consistently achieve performance improvements compared to their respective baseline systems. These findings demonstrate the effectiveness of CFKD and indicate that the proposed framework facilitates more effective utilization of speaker-discriminative information from discrete representations. }

{
Specifically, for ECAPA-TDNN1024, E6 reduces the EER from 3.38 to 2.25 and the minDCF from 0.366 to 0.231, corresponding to a relative performance improvement of approximately 35.3\% compared to E3. Similarly, for ResNet34, E12 decreases the EER from 7.55 to 4.03 and the minDCF from 0.669 to 0.408, yielding a relative improvement of about 42.9\% over E9. The best performance in both architectures is achieved when $\lambda = 40$.} 

It is worth noting that the optimal $\lambda = 40$ is substantially larger than the conventional range ($\lambda < 1$) typically reported in homogeneous distillation settings~\cite{wang2019knowledge}. We hypothesize that this phenomenon arises from the cross-feature nature of our distillation framework. {Since highly compressed token representations entangle speaker identity with other acoustic variations, stronger supervision from the teacher is required to effectively transfer embedding geometry and stabilize optimization. Consequently, a larger distillation weight becomes beneficial in guiding the token-based model toward a well-structured speaker embedding space.
}

{
Notably, model E6 outperforms E2 in Table~\ref{tab:paradox}, confirming the effectiveness of CFKD and indicating that discrete codec tokens preserve sufficient speaker-discriminative information. 
}

\subsection{{Information Compensation from Tokens}}
\label{ssec:error_analysis}

\begin{table}[htbp]
\centering
\caption{Top-5 distinct error intersections among the Teacher and Student models with comparable performance ($\lambda \in \{20, 40, 80\}$). A cross mark ($\times$) denotes that the corresponding model misclassified all trials within that specific subset, while a checkmark ($\checkmark$) indicates that the model classified them correctly.}
\label{tab:error_intersection}
\resizebox{0.9\linewidth}{!}{
\begin{tabular}{c c c c c}
\toprule
\multicolumn{4}{c}{\textbf{Model Involved}} & \textbf{Number of} \\
\cmidrule(lr){1-4}
\textbf{E5 ($\lambda$=20)} & \textbf{E6 ($\lambda$=40)} & \textbf{E7 ($\lambda$=80)} & \textbf{E1 (Teacher)} & \textbf{Error Trials} \\
\midrule
$\times$ & $\times$ & $\times$ & $\times$ & 534 \\ 
$\checkmark$ & $\checkmark$ & $\checkmark$ & $\times$ & 134 \\ 
$\times$ & $\times$ & $\times$ & $\checkmark$ & 114 \\ 
$\times$ & $\checkmark$ & $\checkmark$ & $\checkmark$ & 68 \\  
$\checkmark$ & $\times$ & $\times$ & $\times$ & 51 \\  
\bottomrule
\end{tabular}
}
\end{table}

To further investigate the mechanism behind the performance gains, we list part of intersection of classification errors between the Teacher and Students in Table~\ref{tab:error_intersection}. This analysis reveals insights regarding model capability and feature heterogeneity.

The largest subset (534 trials) consists of hard trials where all models fail. Beyond these common failures, the non-overlapping error sets provide crucial insights into the models' distinct capabilities. Notably, the second largest subset (134 trials) corresponds to errors committed only by the teacher, suggesting that Fbank features possess inherent ``blind spots" or biases. Conversely, the third largest subset (114 trials) comprises cases where all token-based students fail but the teacher succeeds. This shared limitation likely stems from acoustic details being discarded after compression. The prominence of these mutually exclusive error sets explains how the token-based students can approach teacher's performance: the complementary discriminative information captured by the discrete tokens successfully offsets the information loss caused by neural compression.

\subsection{Structural Probing: Inductive Bias and Adaptability}
\label{ssec:inductive_bias}

\label{ssec:inductive_bias}
While both 1D and 2D architectures perform competitively on continuous Fbank features (E1 vs. E8), a stark contrast emerges in the discrete token domain. As shown in Table~\ref{tab:main_results_rel}, ECAPA-TDNN (1D) consistently outperforms ResNet34 (2D) on EnCodec tokens, regardless of distillation. 

To investigate the compatibility between architectural inductive biases and the latent representations of neural codecs, we conduct a \textit{Feature Shuffling} probe. By applying a fixed permutation to the feature dimension, we isolate the impact of feature-dimensional ordering without altering information integrity.

\begin{table}[htbp]
 \caption{Impact of Feature Shuffling on distinct architectures. \textbf{Shuffled} denotes fixed permutation of the feature dimension (pseudo-frequency/channel axis) in model training and test. Models are trained on Vox1 dev set and evaluated on Vox1-O test set.}
 \label{tab:shuffling_analysis}
 \centering
 \resizebox{\linewidth}{!}{ 
 \begin{tabular}{lccccc}
  \toprule
  Backbone & ID & Input & Ordering & EER (\%) & minDCF \\
  \midrule
  \multirow{4}{*}{ECAPA-TDNN1024} 
  & E1 & Fbank & Original & 2.21 & 0.236 \\
  & E14 & Fbank & Shuffled & 2.15 & 0.234 \\
  \cmidrule(lr){2-6}
  & E3 & Tokens & Original & 3.38 & 0.366 \\
  & E15 & Tokens & Shuffled & 3.29 & 0.374 \\
  \midrule
  \multirow{4}{*}{ResNet34} 
  & E8 & Fbank & Original & 2.58 & 0.275 \\
  & E16 & Fbank & Shuffled & 4.96 & 0.508 \\
  \cmidrule(lr){2-6}
  & E9 & Tokens & Original & 7.55 & 0.669 \\
  & E17 & Tokens & Shuffled & 7.45 & 0.678 \\
  \bottomrule
 \end{tabular}
 } 
\end{table}
Table~\ref{tab:shuffling_analysis} reveals a fundamental dichotomy. On Fbanks, ECAPA-TDNN exhibits permutation invariance (E1 $\approx$ E14), whereas ResNet suffers a catastrophic collapse (EER doubles in E16), confirming its reliance on the strong inductive bias of spectral stationarity. Crucially, on tokens, ResNet exhibits poor baseline performance (E9) that remains virtually unchanged after shuffling (E17). This ``invariance to destruction'' empirically shows that the EnCodec latent space is decorrelated regarding spatial adjacency. Consequently, 2D kernels struggle to extract meaningful patterns from discrete tokens. This suggests that order-agnostic architectures like 1D-TDNNs, which treat dimensions as independent channels, are more suitable for processing discrete neural representations.

\subsection{{Experiments on Large Scale Voxceleb2}}

Finally, to validate the robustness of our method, we replicated the experiments on the larger VoxCeleb2 dataset.  Table~\ref{tab:bandwidth_comparison_final_v2} details the results across various bandwidth settings. Compared to the results reported in~\cite{lin2025codec}, our approach demonstrates significant performance improvements across tested bitrates. Notably, at the maximum bitrate, our model achieves a 49.5\% relative improvement (reducing the EER from 2.08\% to 1.05\%), closely approaching the performance of the teacher model.

\section{Conclusion}

We have introduced the Cross-Feature Knowledge Distillation (CFKD) framework to tackle the performance degradation when directly applying neural codec tokens to ASV. CFKD successfully extract speaker identity cues from compressed tokens by leveraging the guidance of a discriminative teacher. Additionally, we empirically demonstrated that the inherent lack of feature-dimensional adjacency makes order-agnostic 1D architectures better suited for token modeling than 2D models. Experiment on VoxCeleb2 dataset confirms that our method is robust across various bitrates, establishing a effective paradigm that significantly narrows the gap between codec-based ASV and continuous feature baselines. 

\section{Acknowledgments}
This work was supported in part by the Research Grants Council of the Hong Kong SAR (Grant No. 15228223), and The Hong Kong Polytechnic University, Project ID P0049192.

\section{Generative AI Use Disclosure}
During the preparation of this work, the authors used Generative AI to improve the readability and language quality of the manuscript, as well as for assistance with \LaTeX\ code formatting. The authors reviewed and edited the output as needed and take full responsibility for the content of the work.
\bibliographystyle{IEEEtran}
\bibliography{mybib}

\end{document}